\documentclass[twocolumn,epsf,letter]{jpsj2}
\title{Novel Pressure Phase Diagram of Heavy Fermion Superconductor CePt$_{3}$Si
\\Investigated by  ac  Calorimetry\thanks{This paper will be published in the July issue of J. Phys. Soc. Jpn.}}

\author{Naoyuki \textsc{Tateiwa}$^{1}$\thanks{E-mail address: tateiwa@popsvr.tokai.jaeri.go.jp}, 
Yoshinori \textsc{Haga}$^{1}$, Tatsuma D. \textsc{Matsuda}$^{1}$, Shugo \textsc{Ikeda}$^{1,2}$, 
\\Takashi \textsc{Yasuda}$^{2}$, Tetsuya \textsc{Takeuchi}$^{3}$, 
Rikio \textsc{Settai}$^{2}$ and Yoshichika \textsc{\=Onuki}$^{1,2}$}

\inst{$^{1}$Advanced Science Research Center, Japan Atomic Energy Research Institute, Tokai, 
Ibaraki 319-1195\\
$^{2}$Department of Physics, Graduate School of Science, Osaka University, Toyonaka, Osaka 560-0043\\
$^{3}$Low Temperature Center, Osaka University, Toyonaka, Osaka 560-0043\\}

\abst{The pressure dependences of the antiferromagnetic and superconducting transition temperatures have been investigated by ac heat capacity measurement under high pressures for the heavy-fermion superconductor CePt$_3$Si without inversion symmetry in the tetragonal structure. The N\'{e}el  temperature $T_{\rm N}$  = 2.2 K decreases with increasing pressure and becomes zero at the critical pressure $P_{\rm AF}$ $\simeq$ 0.6 GPa. On the other hand,  the superconducting phase exists in a wider pressure region from ambient pressure to about 1.5 GPa. The pressure phase diagram of CePt$_3$Si is thus very unique and has never been reported before for other heavy fermion superconductors.}
\kword{heavy-fermion superconductor, CePt$_{3}$Si, ac calorimetry}

\begin{document}
\maketitle

   Recently, Bauer {\it et al.} reported superconductivity in CePt$_3$Si with a non-centrosymmetric tetragonal structure (space group $P$4 $mm$)~\cite{bauer}. 
Superconductivity with the transition temperature $T_{\rm sc}$  = 0.75 K is
realized in the long-range antiferromagnetic ordered state with the N\'{e}el temperature $T_{\rm N}$  = 2.2 K. A large electronic specific heat coefficient $\gamma$ = 300 - 400 mJ/K$^2{\cdot}$mol and a large slope of upper critical field $dH_{\rm c2}/dT$ (=  -8.5 T/K) at $T_{\rm sc}$ suggest that superconductivity is based on heavy quasiparticles. 
     
 CePt$_3$Si is the first heavy-fermion superconductor lacking the inversion center in the crystal structure~\cite{bauer,saxena}. 
It has been thought that a material lacking inversion symmetry would be an unlikely candidate for spin-triplet pairing~\cite{anderson,sigrist}. Interestingly, the upper critical field $H_{\rm c2}(0)$ ($\sim$ 3 T)  in CePt$_3$Si exceeds the Pauli-Clogston limit ($\sim$ 1 T), and triplet paring was suggested in the superconducting state of CePt$_3$Si~\cite{bauer}. Recent theoretical studies have claimed that the absence of the inversion center does not indiscriminately suppress the spin-triplet
 pairing state~\cite{frigeri1,samokhin1,sergienko,samokhin2,frigeri2,mineev}. It should be noted that recently superconductivity was found in a ferromagnet UIr without inversion symmetry under pressure where the Curie temperature approximately becomes zero~\cite{akazawa}. The relation between anisotropic superconductivity and the non-centrosymmetric crystal structure is the most crucial issue to be clarified at present.
 
  We grew a high-quality single crystal of CePt$_3$Si and investigated magnetic and electrical properties~\cite{takeuchi,hashimoto}.  It was clarified that $H_{\rm c2}(0)$ was approximately isotropic: $H_{\rm c2}(0)$ = 2.7 T for $H$ $\parallel$ [100] and 3.2 T for $H$ $\parallel$ [001].  In the de Haas-van Alphen experiment,  the topology of the Fermi 
surface in CePt$_3$Si was found to be most likely similar to that of LaPt$_3$Si. Large cyclotron masses of 10-20 $m_{\rm 0}$ were detected in CePt$_3$Si, which indicates the existence of heavy quasiparticles in this compound.  In the neutron scattering experiment, a clear antiferromagnetic Bragg peak with $Q$ = (0,0,1/2) was observed below $T_{\rm N}$ and the magnetic moment was determined as 0.16 ${\mu}_{\rm B}$/Ce~\cite{metoki}. The Bragg peak intensity was almost constant below $T_{\rm sc}$, which indicates that the antiferromagnetic state coexists with the superconducting 
state.  A microscopic coexistence between magnetism and superconductivity was also suggested by muon spin relaxation (${\mu}$SR) and NMR experiments~\cite{amato,higemoto,yogi}. In particular, in the NMR experiment, the novel behavior of 1/$T_{1}T$ (1/$T_{1}$: $^{195}$Pt nuclear spin-lattice relaxation rate) was observed below $T_{\rm sc}$, suggesting that a new class of superconducting state is realized in CePt$_3$Si.

  Futhermore, the pressure dependence of the superconducting transition temperature $T_{\rm sc}$ was investigated by electrical resistivity measurements~\cite{yasuda,nicklas}. The transition temperature 
decreases with increasing pressure and finally becomes zero at 1.5 GPa. The N\'{e}el temperature $T_{\rm N}$ also decreases with increasing pressure, but  the pressure dependence is 
not clear above 0.8 GPa because the change in the resistivity at $T_{\rm N}$ is too weak to be 
detected. In order to clarify the pressure phase diagram of CePt$_3$Si, we carried out heat 
capacity measurement under high pressures.
    
 A detailed description of our sample preparation is given in the previous papers~\cite{takeuchi,hashimoto,yasuda}. In the present heat capacity measurement, we used a 
 high-quality single-crystal sample, which was cut from the sample used in the previous de Haas-van Alphen experiment. This sample was also used in the pressure experiment for the electrical resistivity and ac susceptibility. The sample size for the ac heat capacity measurement was 0.35${\times}$0.25${\times}$0.05 mm$^3$.

    Heat capacity was measured by the ac calorimetry method in a piston cylinder pressure cell. 
The sample was thermally linked with a heat bath and was heated by application of an oscillating heating power 
$P$ = $P_0$[1+cos(${\omega}$t)] , generated by a current of frequency ${\omega}$/2. The temperature oscillation $T_{\rm ac}$ at the same frequency ${\omega}$ 
was detected with a thermometer attached to the sample. Here, $T_{\rm ac}$ depends on the working frequency ${\omega}$, 
the heat capacity of the sample $C$, and the thermal conductivity $\kappa$ through the following equation~\cite{sullivan,wilhelm} $ T_{\rm ac} = P_{0}/({\kappa}+iC{\omega})$, with $i^2$ = -1, under the assumption that the thermal relaxation in a sample is faster than 1/$\omega$. 
When working above the cut-off frequency ${\omega}_1$ = ${\kappa}/C$, $T_{\rm ac}$ is inversely proportional 
to $C{\omega}$. The temperature oscillation $T_{\rm ac}$ was measured with a AuFe/Au thermocouple 
(Au + 0.07 at \% Fe) with a diameter of 25 ${\mu}$m which was directly bonded to the sample. 
The thermovoltage $V_{\rm ac}$ is induced by the temperature difference between the sample 
at the temperature ($T_0$ + ${\Delta}T$) and the edge of the sample chamber at $T_0$. The value of $T_{\rm ac}$ 
is thus given as $T_{\rm ac}$=$V_{\rm ac}/S$, where $S$ is the thermopower of a thermocouple. The sample was heated with a current supplied through Au wires with a diameter of 10 ${\mu}$m attached to the sample. Frequencies around 100 Hz were used for the measurement. In the present method, we could not determine the absolute value of the heat capacity. The Au-wires and thermocouple contribute to the heat capacity as a background. It is, however, noted that the signal is mainly due to the heat capacity of the sample at low temperatures 
because the volume ratio of the wires to the sample is in the order of a few ${\%}$.  
The low-temperature measurement was performed using a $^{3}$He-$^{4}$He dilution refrigerator. 
We used a hybrid piston cylinder-type cell, where the Co-Ni-Cr-Mo (MP35N) inner cylinder was inserted into the Cu-Be outer sleeve~\cite{walker,uwatoko}. As a pressure transmitting 
medium, Daphne oil (7373) was used.
   
   Figure 1(a) shows the heat capacity $C_{\rm ac}$ under high pressures.  Experimental data obtained under high pressures are shifted downwards.  Here, $C_{\rm ac}$ corresponds to the inverse of the amplitude of the temperature oscillation $T_{\rm ac}$. At ambient pressure, a clear heat  capacity peak was observed at the N\'{e}el temperature $T_{\rm N}$.  The N\'{e}el temperature was determined as $T_{\rm N}$ = 2.2 K, as shown by an arrow. The heat capacity peak around $T_{\rm N}$ is of the ${\lambda}$-type, and the peak structure is sharper than those in the previous reports~\cite{bauer,takeuchi}. This might reflect the high quality of the present sample.
\begin{figure}[t]
\begin{center}
\includegraphics[width=7.8cm]{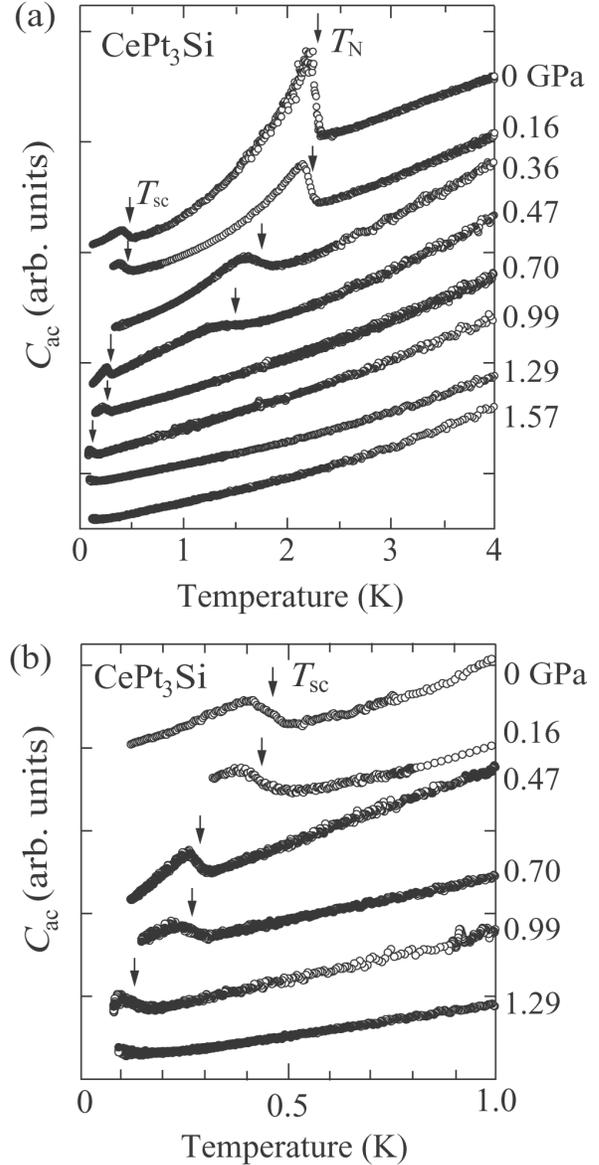}
 \end{center}
\caption{\label{fig:epsart} (a)Temperature dependence of heat capacity $C_{\rm ac}$ and (b) low-temperature part of heat capacity $C_{\rm ac}$ of CePt$_3$Si under high pressures. Experimental data under high pressures are shifted downwards. $T_{\rm N}$ and $T_{\rm sc}$ correspond to the N\'{e}el temperature and superconducting transition temperature, respectively. }
\end{figure} 

 Below 0.5 K, the heat capacity indicates another peak due to the superconducting transition. 
The transition temperature $T_{\rm sc}$  is determined as $T_{\rm sc}$ = 0.46 K. The value is lower than those determined by the resistivity and ac magnetic susceptibility measurements ($\sim$ 0.7 K)~\cite{takeuchi}. The reason for this discrepancy is not clear.  Such a discrepancy was also found in the heavy fermion superconductor CeIrIn$_5$~\cite{petrovic}.  

 The value of $\it{\Delta{C_{\rm ac}}/C_{\rm ac}(T_{\rm sc})}$ is 0.33 at ambient pressure. Here, $\it{\Delta{C_{\rm ac}}}$ is the jump of the heat capacity at $T_{\rm sc}$ and $C_{\rm ac}(T_{\rm sc})$ is the value of $C_{\rm ac}$ just above $T_{\rm sc}$, namely, corresponding to ${\gamma}T_{\rm sc}$, where ${\gamma}$ is the electronic specific heat coefficient. It is noted that a smaller value of 0.25 was reported in ref. 1. These values are far smaller than the BCS value ${\it{\Delta}}C/({\gamma}T_{\rm sc})$ = 1.43. This reduction might be due to the gapless structure of the anisotropic superconducting gap and/or the high sensitivity of the present superconductivity to impurities.

 With increasing pressure, the antiferromagnetic ordering 
shifts to lower temperatures. The peak at  $T_{\rm N}$  = 2.2 K becomes 
weak with increasing pressure and finally becomes a broad hump at 0.47 GPa.  
The peak was not observed in the $C_{\rm ac}$ curve at 0.70 GPa. 
This indicates that the antiferromagnetic ordering disappears at 0.70 GPa. 
The antiferromagnetic critical pressure $P_{\rm AF}$ was thus estimated as $P_{\rm AF}$ $\simeq$ 0.6 GPa. 

  Figure 1(b) shows the low-temperature part of $C_{\rm ac}$.  The peak structure appears at $T_{\rm sc}$ under pressure. $T_{\rm sc}$ decreases with increasing pressure. The peak structure becomes weak at higher pressures. There is, however,  no peak structure at 1.29 and 1.57 GPa (not shown in Fig. 1 (b)). The present result is approximately consistent with that of our previous resistivity measurement under high pressures; the superconducting critical pressure $P_{\rm sc}$ was estimated as $P_{\rm sc}$ $\simeq$ 1.5 GPa~\cite{yasuda}. The $T_{\rm sc}$ value at 1.29 GPa might be below the lowest temperature of our measurement, 90 mK. In fact, $C_{\rm ac}$ shows an upturn structure below 130 mK at 1.29 GPa. 
\begin{figure}[t]
\begin{center}
\includegraphics[width=8cm]{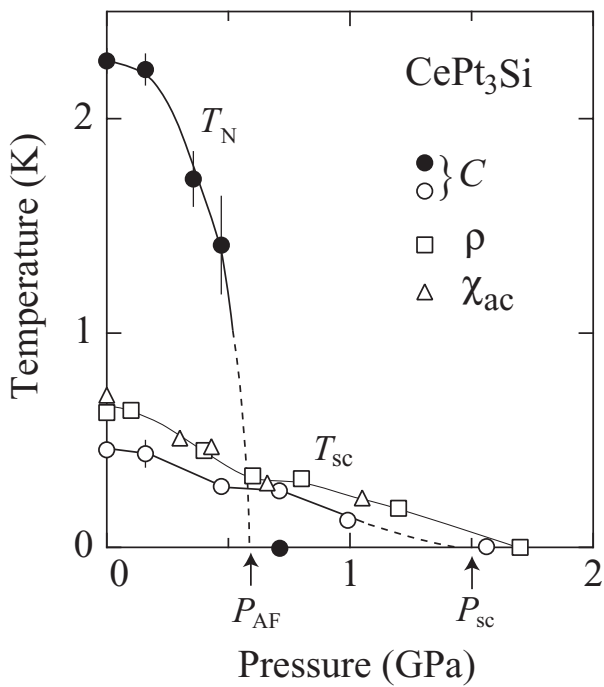}
 \end{center}
\caption{\label{fig:epsart} Pressure phase diagram of CePt$_3$Si. The data shown by closed and open circles correspond to $T_{\rm N}$ and $T_{\rm sc}$, respectively. The value of $T_{\rm sc}$ determined by the resistivity and ac susecptibility measurements are shown by open squares and triangles, respectively.  Solid and dotted lines are guides to the eyes.}
\end{figure}  
   
  Figure 2 shows the pressure phase diagram of CePt$_3$Si determined by 
the present heat capacity measurement, together with the superconducting  transition temperatures determined by our previous resistivity $\rho$ and ac susceptibility $\chi_{\rm ac}$ measurements~\cite{yasuda}. We did not adopt the N\'{e}el  temperature determined by the resistivity measurement in Fig. 2 because the change in the resistivity at $T_{\rm N}$ is very small. The $T_{\rm N}$ value decreases faster than that of $T_{\rm sc}$ with increasing pressure. An interesting point is that the superconducting phase exists above $P_{\rm AF}$. 

 The pressure dependence of $T_{\rm sc}$ is characteristic, as shown in Fig. 2. Namely, $T_{\rm sc}$ decreases as a function of pressure, becomes approximately constant from 0.6 GPa to 0.8 GPa, decreases further with increasing pressure, and becomes zero at $P_{\rm sc}$ $\simeq$ 1.5 GPa.  The present pressure of 0.6 GPa corresponds to the antiferromagnetic critical pressure $P_{\rm AF}$ $\simeq$ 0.6 GPa.  In the pressure region from $P_{\rm AF}$ $\simeq$ 0.6 GPa to 1.5 GPa, CePt$_3$Si is in the paramagnetic state and shows only the superconducting transition. The pressure phase diagram of CePt$_3$Si is thus very unique.
 
  Figure 3(a) shows the pressure dependence of $\it{\Delta{C_{\rm ac}}/C_{\rm ac}(T_{\rm sc})}$. It is almost constant in the pressure region from ambient pressure to $P_{\rm AF}$ and decreases steeply above $P_{\rm AF}$ $\simeq$ 0.6 GPa. The value of $\it{\Delta{C_{\rm ac}}/C_{\rm ac}(T_{\rm sc})}$ is, however, about 0.2 at 0.99 GPa. This indicates that the bulk 
superconducting state survives in the paramagnetic phase, namely above $P_{\rm AF}$ $\simeq$ 0.6 GPa.

   \begin{figure}[t]
\begin{center}
\includegraphics[width=7.8cm]{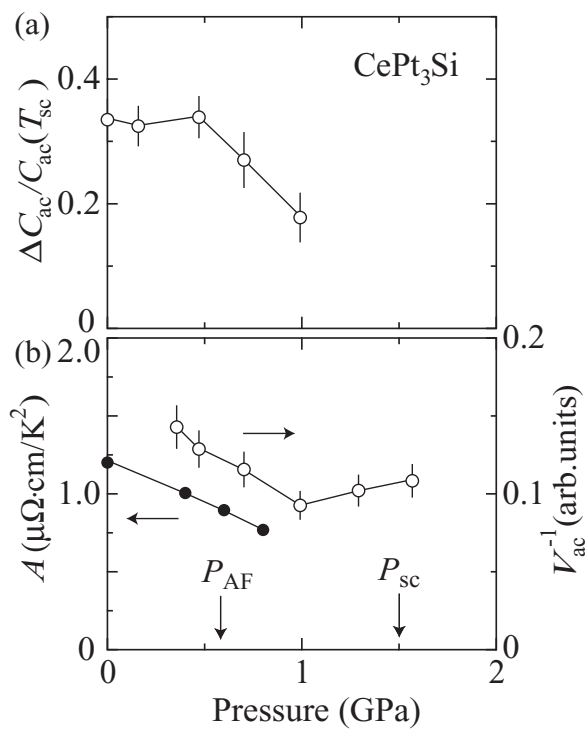}
 \end{center}
\caption{\label{fig:epsart} (a) Pressure dependence of the jump of the heat capacity of CePt$_3$Si at  $T_{sc}$. (b) Pressure dependences of the coefficient of the $T^2$ term in the resistivity (left side) and ${V_{ac}^{-1}}$ (right side) for CePt$_3$Si.}
\end{figure}

   The thermocouple AuFe/Au is a dilute Kondo system and the thermopower $S(T)$ is given as $S(T)$ $\propto$ $T$ below 1 K.  Therefore, the value of the inverse thermovoltage ${V_{\rm ac}^{-1}}$, which corresponds to $[S(T)T_{\rm ac}]^{-1}$, is approximately proportional to $C/T$ at low temperatures. Therefore, the pressure dependence of $V_{\rm ac}^{-1}$ roughly corresponds to the electronic specific heat coefficient $\gamma$.  
   
   Figure 3(b) shows the pressure dependence of ${V_{\rm ac}^{-1}}$, together with that of the coefficient $A$ of the $T^{2}$ term in the electrical resistivity. Here, the data ${V_{\rm ac}^{-1}}$ was obtained just above $T_{\rm sc}$ under the same conditions of heater power $P_{0}$,  ac current frequency $\omega$, and thermal conductivity $\kappa$. At 1.29 and 1.57 GPa, where the superconducting transition does not appear, ${V_{ac}^{-1}}$ becomes almost constant below 0.2 K. This constant value is plotted in Fig. 3. 
The value of ${V_{\rm ac}^{-1}}$ decreases with increasing pressure and shows a tendency to saturate above 1.0 GPa, as shown in Fig. 3(b). Similarly, the $A$ value simply decreases with increasing pressure and there is no anomaly around $P_{\rm AF}$. 
These results indicate that there is no divergent feature in the pressure dependence of the electronic specific heat coefficient $\gamma$ at $P_{\rm AF}$ and most likely at $P_{\rm sc}$.   
   
  We will compare the present result for CePt$_3$Si with those for the other heavy-fermion Ce-based superconductors. In a prototype superconductor  CeCu$_2$Si$_2$, superconductivity occurs in a  nonmagnetic state that is close to antiferromagnetic instability~\cite{steglich1,steglich2}.  The electronic state can be tuned by pressure or the stoichiometric composition of a sample. 
Superconductivity and antiferromagnetism are basically competitive and do not coexist in the compound. This is quite different from the characteristic feature of CePt$_3$Si. Namely, superconductivity coexists with antiferromagnetism in CePt$_3$Si.

 Another example is pressure-induced superconductivity in CeIn$_3$, CeRh$_2$Si$_2$, and CePd$_2$Si$_2$~\cite{mathur,knebel,movshovich,araki,demuer}. These compounds show antiferromagnetic ordering at ambient pressure.  $T_{\rm N}$ decreases with increasing pressure. Superconductivity appears around the magnetic critical region where $T_{\rm N}$ becomes zero, namely around $P_{\rm AF}$. Superconductivity is considered to be mediated by low-energy magnetic excitations around the magnetic critical region where the heavy-fermion state is realized. The superconducting phase exists in a narrow pressure region around the magnetic critical region, and $T_{\rm sc}$ becomes a maximum around the critical pressure $P_{\rm AF}$. These features are quite different from those of CePt$_3$Si where the bulk superconducting phase exists in a wide pressure region above and below $P_{\rm AF}$, and $T_{\rm sc}$ does not shows a maximum at $P_{\rm AF}$. The maximum $T_{\rm sc}$ is realized at ambient pressure. The present superconductivity is most robust at ambient pressure. It is suggested from the pressure dependence of  $V_{\rm ac}$ that the critical pressure $P_{\rm AF}$ is not of second order but of first one. Therefore, superconductivity in CePt$_3$Si is different from superconductivity associated with magnetic instability around the magnetic critical region. 
 
  The relation between antiferromagnetism and superconductivity in CePt$_3$Si is thus very unique. One might speculate that the superconducting and antiferromagnetic phases compete with each other below $P_{\rm AF}$ and the former overcomes the latter above $P_{\rm AF}$. However, it should be noted that both ordering temperatures decrease with increasing pressure up to $P_{\rm AF}$ and that both $T_{\rm sc}$ and $\it{\Delta{C_{\rm ac}}/C_{\rm ac}(T_{\rm sc})}$ decrease above $P_{\rm AF}$. It seems to be difficult to consider these experimental results from a competitive relation between the two states. From the pressure dependence of $T_{\rm sc}$, it is supposed that the coupling between the antiferromagnetic and superconducting states is basically weak. In order to clarify the relation between the two states, further study is needed. In particular, it is needed to investigate the change in the microscopic superconducting properties such as the paring symmetry across $P_{\rm AF}$.

  In conclusion, we constructed the pressure phase diagram of the heavy-fermion superconductor CePt$_3$Si by ac calorimetry. The bulk superconducting phase 
exists in a wide pressure region from ambient pressure to about 1.5 GPa, which is far above the antiferromagnetic critical pressure $P_{\rm AF}$ $\simeq$ 0.6 GPa. The overall features of pressure phase diagram of CePt$_3$Si are different from those of the other heavy-fermion superconductors. Furthermore, it is emphasized, on the basis of the results of the present pressure experiment, that superconductivity in CePt$_3$Si is most robust at ambient pressure where the antiferromagnetic ordered state is realized.

\section*{Acknowledgments} 
 One of authors (N.T) thanks Dr. M. -A. M\'{e}asson for helpful discussions. This work was financially supported by Grants-in-Aid for Young Scientists (B), Scientific Research (A), Creative Research (15GS0213), Scientific Research in Priority Area "Sukutterdite" (No. 16037215) from the Ministry of Education, Culture, Sports, Science and Technology (Mext) and the Japanese Society for the Promotion of Science.

\end{document}